\documentclass[fleqn,10pt]{wlscirep}
\usepackage[utf8]{inputenc}
\usepackage[T1]{fontenc}
\usepackage{bm}
\title{Firefly swarms: What models for what physics?}

\author[]{Rapha\"el Sarfati}
\affil[]{Cornell University, Ithaca, NY, USA \\ raphael.sarfati@cornell.edu}

\begin{abstract}
The following is a response to \href{https://www.nature.com/articles/s42254-023-00675-z}{\textit{A new chapter in the physics of firefly swarms}}, a 2024 \href{https://www.nature.com/natrevphys/for-authors/preparing-your-submission}{Comment} in Nature Reviews Physics.
I contend that the (non peer-reviewed) Comment is generally imprecise and factually inaccurate.
Following the journal's guidelines, I wrote and submitted this 500-word \href{https://www.nature.com/documents/natrev-articleformatguide-Correspondence.pdf}{Correspondence}.
It was promptly rejected by the editors, partly on the basis that the original Comment is merely an ``opinion piece''.
I provide here my draft as submitted.
Further details are available upon request.
\end{abstract}
\begin{document}

\flushbottom
\maketitle

\thispagestyle{empty}
\vspace{-0.7cm}
\paragraph{}
In a recent Comment (Peleg, O. A new chapter in the physics of firefly swarms, Nat. Rev. Phys. 2024), Orit Peleg calls for ``context-specific mathematical models'' to ``deepen our understanding of the physics of firefly swarms''~\cite{Peleg2024}. 
Despite reviewing recent publications, the Comment fails to define what a `physics of firefly swarms' should entail and what new models should explain. 
I propose some insights.  
 
\paragraph{}
Animal collective behavior produces fascinating dynamics. 
Physicists find analogies between bird flocks and spin systems~\cite{Bialek2012}, fish schools and thermodynamics~\cite{Giannini2020}, insect swarms and self-gravitating systems~\cite{Gorbonos2020}. 
The underlying question is often that of emergence: how do we go from local interactions between elementary constituents to spontaneous, ensemble patterns with specific statistical signatures~\cite{Ouellette2022}? 
 
\paragraph{}
Accordingly, what would a firefly physicist investigate? 
While a broad question, possible answers follow. 
One could seek patterns in the collective flashing of fireflies; propose equations that generate these patterns; and uncover commonalities with other systems. 
Indeed, modern physics is often concerned with universality~\cite{Parisi2023}.  
 
\paragraph{}
Hence, we studied firefly flash synchronization because: 1) an emergent pattern is apparent; and 2) synchrony is ubiquitous~\cite{Strogatz2003}. 
The latter prompted mathematicians, decades ago, to put synchronization into equations. 
They showed that coupled oscillators usually converge towards a common phase.  
 
\paragraph{}
Foundational models were not concerned with fireflies specifically, but rather a general understanding of synchrony. 
Unsurprisingly, when we collected data from natural swarms~\cite{Sarfati2023}, we realized that not only the microscopic details (fireflies are not continuously coupled oscillators), but also the macroscopic output of abstract models departed from empirical observations.  
 
\paragraph{}
For example, \textit{P. carolinus} fireflies synchronize during flash bursts which become periodic only in large swarms. 
The emergent periodicity was explained by a simple paradigm proposed by Iyer-Biswas and Joshi: individual bursts occur at random, but trigger all other fireflies, resulting in regression to the smallest interburst duration~\cite{Sarfati2023}. 
 
\paragraph{}
The fact that a simple principle, rather than convoluted equations, resolved the puzzle is telling. Simple principles can be insightful, while sophisticated models might overfit, rather than elucidate, the data. 
 
\paragraph{}
The utility and purpose of a mathematical model is a vast question, discussed in the context of collective behavior notably by Ouellette~\cite{Ouellette2015}. 
A specialized model could bridge the gap between microscopic details and macroscopic phenomenology. Alternatively, a unifying model could reveal commonalities across systems by distilling their essential physics, regardless of specificities. 
In this spirit, Cavagna \textit{et al.} make an elegant case for the need to `keep it simple' as they construct a physics of flocking~\cite{Cavagna2018}. 
 
\paragraph{}
In any case, new models should follow well-defined objectives. Is there a practical necessity for finer equations of collective flashing? The Comment suggests firefly conservation, which, absent further elaboration, seems unconvincing. Are there unresolved questions about blinking swarms? Certainly, but these ought to be formulated rigorously.  
 
\paragraph{}
One such outstanding question is: why do males synchronize, when they are competing for female attention? Hypothesizing that synchronization solves the `cocktail party problem'~\cite{Peleg2024} seems unfounded: variability, rather than homogeneity, enables identifiability (try singling out just one voice amidst a choir!).  
 
\paragraph{}
Although, perhaps, the answer to `why' lies beyond the purview of physics...

\bibliography{firefly-correspondence}

\begin{thebibliography}{10}
\urlstyle{rm}
\expandafter\ifx\csname url\endcsname\relax
  \def\url#1{\texttt{#1}}\fi
\expandafter\ifx\csname urlprefix\endcsname\relax\def\urlprefix{URL }\fi
\expandafter\ifx\csname doiprefix\endcsname\relax\def\doiprefix{DOI: }\fi
\providecommand{\bibinfo}[2]{#2}
\providecommand{\eprint}[2][]{\url{#2}}

\bibitem{Peleg2024}
\bibinfo{author}{Peleg, O.}
\newblock \bibinfo{journal}{\bibinfo{title}{A new chapter in the physics of firefly swarms}}.
\newblock {\emph{\JournalTitle{Nature Reviews Physics}}} \textbf{\bibinfo{volume}{6}}, \bibinfo{pages}{72--74}, \doiprefix\url{https://doi.org/10.1038/s42254-023-00675-z} (\bibinfo{year}{2024}).

\bibitem{Bialek2012}
\bibinfo{author}{Bialek, W.} \emph{et~al.}
\newblock \bibinfo{journal}{\bibinfo{title}{Statistical mechanics for natural flocks of birds}}.
\newblock {\emph{\JournalTitle{Proceedings of the National Academy of Sciences}}} \textbf{\bibinfo{volume}{109}}, \bibinfo{pages}{4786--4791}, \doiprefix\url{https://doi.org/10.1073/pnas.1118633109} (\bibinfo{year}{2012}).

\bibitem{Giannini2020}
\bibinfo{author}{Giannini, J.~A.} \& \bibinfo{author}{Puckett, J.~G.}
\newblock \bibinfo{journal}{\bibinfo{title}{Testing a thermodynamic approach to collective animal behavior in laboratory fish schools}}.
\newblock {\emph{\JournalTitle{Phys. Rev. E}}} \textbf{\bibinfo{volume}{101}}, \bibinfo{pages}{062605}, \doiprefix\url{http://doi.org/10.1103/PhysRevE.101.062605} (\bibinfo{year}{2020}).

\bibitem{Gorbonos2020}
\bibinfo{author}{Gorbonos, D.} \emph{et~al.}
\newblock \bibinfo{journal}{\bibinfo{title}{Similarities between insect swarms and isothermal globular clusters}}.
\newblock {\emph{\JournalTitle{Phys. Rev. Res.}}} \textbf{\bibinfo{volume}{2}}, \bibinfo{pages}{013271}, \doiprefix\url{https://doi.org/10.1103/PhysRevResearch.2.013271} (\bibinfo{year}{2020}).

\bibitem{Ouellette2022}
\bibinfo{author}{Ouellette, N.~T.}
\newblock \bibinfo{journal}{\bibinfo{title}{A physics perspective on collective animal behavior}}.
\newblock {\emph{\JournalTitle{Physical Biology}}} \textbf{\bibinfo{volume}{19}}, \bibinfo{pages}{021004}, \doiprefix\url{https://doi.org/10.1088/1478-3975/ac4bef} (\bibinfo{year}{2022}).

\bibitem{Parisi2023}
\bibinfo{author}{Parisi, G.}
\newblock \emph{\bibinfo{title}{In a Flight of Starlings}} (\bibinfo{publisher}{Penguin Random House}, \bibinfo{address}{New York}, \bibinfo{year}{2023}).

\bibitem{Strogatz2003}
\bibinfo{author}{Strogatz, S.~H.}
\newblock \emph{\bibinfo{title}{Sync: How Order Emerges from Chaos in the Universe, Nature, and Daily Life}} (\bibinfo{publisher}{Hyperion}, \bibinfo{address}{New York}, \bibinfo{year}{2003}).

\bibitem{Sarfati2023}
\bibinfo{author}{Sarfati, R.} \emph{et~al.}
\newblock \bibinfo{journal}{\bibinfo{title}{Emergent periodicity in the collective synchronous flashing of fireflies}}.
\newblock {\emph{\JournalTitle{eLife}}} \textbf{\bibinfo{volume}{12}}, \bibinfo{pages}{e78908}, \doiprefix\url{https://doi.org/10.7554/eLife.78908} (\bibinfo{year}{2023}).

\bibitem{Ouellette2015}
\bibinfo{author}{Ouellette, N.~T.}
\newblock \bibinfo{journal}{\bibinfo{title}{Empirical questions for collective-behaviour modelling}}.
\newblock {\emph{\JournalTitle{Pramana}}} \textbf{\bibinfo{volume}{84}}, \bibinfo{pages}{353--363}, \doiprefix\url{https://doi.org/10.1007/s12043-015-0936-5} (\bibinfo{year}{2015}).

\bibitem{Cavagna2018}
\bibinfo{author}{Cavagna, A.}, \bibinfo{author}{Giardina, I.} \& \bibinfo{author}{Grigera, T.~S.}
\newblock \bibinfo{journal}{\bibinfo{title}{The physics of flocking: Correlation as a compass from experiments to theory}}.
\newblock {\emph{\JournalTitle{Physics Reports}}} \textbf{\bibinfo{volume}{728}}, \bibinfo{pages}{1--62}, \doiprefix\url{https://doi.org/10.1016/j.physrep.2017.11.003} (\bibinfo{year}{2018}).

\end{thebibliography}

\section*{Acknowledgments}
I thank colleagues for insightful conversations on this topic.


\section*{Competing interests}
The author declares no competing interests. 

\section*{Note}
The author used to work as a postdoc under Orit Peleg's supervision.



\end{document}